# High computational density nanophotonic media for machine learning inference

Zhenyu Zhao[1], Yichen Pan[1], Jinlong Xiang[1], Yujia Zhang[1], An He[1], Yaotian Zhao[1], Youlve Chen[1], Yu He[1], Xinyuan Fang[2,*], Yikai Su[1], Min Gu[2,] and Xuhan Guo[1,*]

[1]State Key Laboratory of Advanced Optical Communication Systems and Networks, Department of Electronic Engineering, Shanghai Jiao Tong University; Shanghai, 200240, China.

[2]School of Artificial Intelligence Science and Technology, University of Shanghai for Science and Technology; Shanghai, 200093, China.

*Corresponding author. Email: xinyuan.fang@usst.edu.cn, guoxuhan@sjtu.edu.cn.

**Abstract**

**Efficient machine learning inference is essential for the rapid adoption of artificial intelligence (AI) across various domains. On-chip optical computing has emerged as a transformative solution for accelerating machine learning tasks, owing to its ultra-low power consumption. However, enhancing the computational density of on-chip optical systems remains a significant challenge, primarily due to the difficulties in miniaturizing and integrating key optical interference components. In this work, we harness the potential of fabrication-constrained scattering optical computing within nanophotonic media to address these limitations. Central to our approach is the use of fabrication-aware inverse design techniques, which enable the realization of manufacturable on-chip scattering structures under practical constraints. This results in an ultra-compact optical neural computing architecture with an area of just 64 μm²—representing a remarkable three orders of magnitude reduction in footprint compared to traditional optical neural networks. Our prototype, tested on the Iris flower dataset, achieved an experimental accuracy of 86.7%, closely matching the simulation benchmark. This breakthrough showcases a promising pathway toward ultra-dense, energy-efficient optical processors for scalable machine learning inference, significantly reducing both the hardware footprint, latency, and power consumption of next-generation AI applications.**

Introduction

Large-scale machine learning models are driving transformative advances across various industries, with examples like generative pre-trained transformers revolutionizing natural language processing[1] and vision transformers enhancing weather forecasting[2]. However, these breakthroughs come with a significant cost: a substantial increase in computational power demands. Traditional electronic computing platforms are reaching their limits in power efficiency, as further reductions in transistor energy consumption yield diminishing returns. As a result, integrated optical computing platforms, which offer reduced power consumption and inherent parallelism, are garnering increasing research attention[3].

Integrated optical architectures for machine learning, such as Mach–Zehnder interferometer (MZI) meshes[4,5], waveguide attenuators[6,7], micro-rings[8], and diffractive elements[9–12], have been extensively explored as power-efficient solutions for optical neural computing. However, these architectures rely on systematically arranged units for phase and attenuation tuning, which are constrained by the empirical design principles of their constituent components, thus limiting further miniaturization. Recently, machine learning using nanophotonic media has emerged as a promising approach to reducing on-chip optical computation footprints by leveraging first-principles design[13,14], with similar approaches also applied to on-chip matrix operations[15,16]. Nevertheless, the construction of compact and functional neural computing structures using nanophotonic media that seamlessly integrate fabrication constraints into the training process remains an experimental challenge. Developing such structures compatible with general fabrication technologies is eagerly pursued, as it holds the potential to unlock the full capabilities of integrated photonic platforms for machine learning applications.

In this work, we present a novel approach to on-chip optical computing, leveraging nanophotonic media to address the challenges of computational density and power efficiency in machine learning inference. For the first time, we demonstrate a fully fabricated and experimentally validated neural computing structure on a silicon-on-insulator (SOI) platform using nanophotonic media, tested with the Iris flower dataset[17]. Our design achieved an unprecedentedly compact footprint of just 64 µm²—orders of magnitude smaller than traditional on-chip optical neural networks, which typically occupy square millimeter scales. After training, the model achieved a test accuracy of

86.7%, with the experimental accuracy of the fabricated chip also reaching 86.7%. To achieve this, we developed a novel training method that incorporates fabrication constraints directly into the training process, coupled with low-index-contrast structures for enhancing fabrication tolerance. These innovations not only significantly reduce the device footprint but also offer scalability to more complex machine learning tasks. To further demonstrate the scalability of our approach, we applied it to a more complex dataset for handwritten digit image recognition, achieving a test accuracy of 92.8%. This work paves the way for ultra-compact, power-efficient optical neural computing systems, significantly reducing the inference costs of large-scale machine learning models and providing a clear path toward scalable optical AI hardware.

**Results**

**On-Chip Nanophotonic Media Configuration**

On-chip optical devices can be conceptualized as systems that perform operations on a set of basic optical modes[18]. Similarly, on-chip analog computing systems process input information encoded in these basic modes and transform it into output modes carrying the desired information. The efficiency of this transformation, commonly referred to as coupling efficiency, directly determines the computational density of the on-chip optical computing system. As illustrated in Fig. **1a**, coupling efficiency (κ), as predicted by coupled mode theory[19], is influenced by refractive index perturbations within the system. Specifically, larger perturbations and greater refractive index contrasts result in higher coupling efficiencies. Consequently, enhancing these two factors is key to improving computational density in on-chip optical computing systems.

Fig. **1b** compares existing structures for on-chip optical computing systems with the proposed nanophotonic media, demonstrating that nanophotonic media can achieve relatively higher average efficiency. Traditional approaches, such as electric tuning of doped waveguides, thermal tuning of waveguides, or etching diffractive structures, rely on explicit design theories. However, these methods are often constrained by empirical design principles, which limit their ability to achieve high

mode coupling efficiency. In contrast, nanophotonic media overcome the limitations of empirical design principles by enabling high coupling efficiency within a compact area, thereby achieving greater computational density in on-chip systems. Nonetheless, the high coupling efficiency of nanophotonic media in such small regions also amplifies the sensitivity to fabrication-induced perturbations, which can introduce significant errors. Therefore, incorporating fabrication constraints into the design process of nanophotonic media is essential to ensure reliable system performance.

Machine learning inference in our approach is achieved using the high computational density nanophotonic media, as shown in Fig. **1c**, where the Iris flower classification task is taken as an example. Features are encoded into the phase modulation of waveguide eigenmodes on the left, while inference results are carried by the optical power in different output waveguides. The inference function is carried out by the nanophotonic media within the central block, where the radii of the holes in the nanophotonic media act as trainable parameters during the training process.

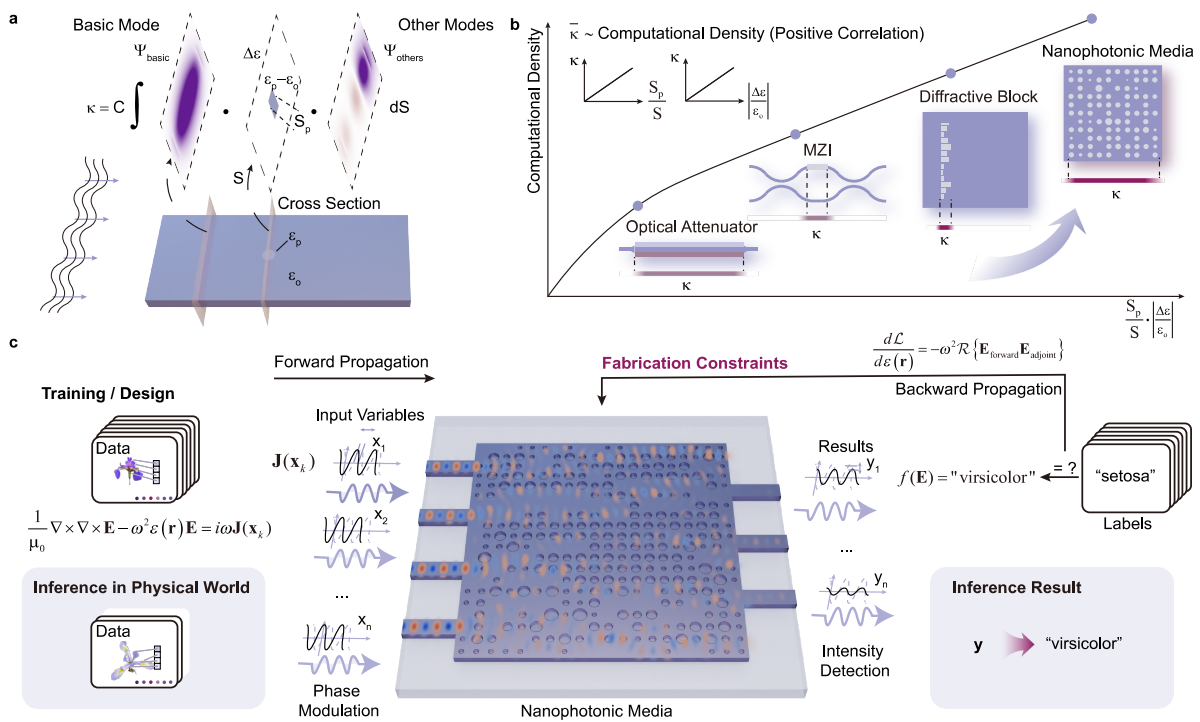

**Fig. 1 | Schematic of design and on-chip machine learning inference with nanophotonic media. a,** qualitative analysis for the coupling efficiency which is related to the computational density in on-chip optical systems. **b,** comparison of different kinds of on-chip optical computational systems with computation-related

coupling efficiency. **c,** schematic of the training and inference process of the nanophotonic medium, where gradient information is calculated via the adjoint method and used to update the structures according to fabrication constraints. After fabrication, inference is achieved by light passing through the scattering nanophotonic medium. Input information is encoded in the phase-modulated input light, and results are obtained through intensity detection at the output ports.

The training process can be viewed as solving an optimization problem under fabrication constraints and physical constraints:

$$\begin{aligned} &\text{minimize} \quad L(\mathbf{T}(\mathbf{r},\mathbf{X}),\mathbf{t}) \\ &\text{subject to} \quad g(r_j,\mathbf{r}) \leq 0, \forall j \\ &\qquad \frac{1}{\mu_0} \nabla \times \nabla \times \mathbf{E} - \omega^2 \varepsilon(\mathbf{r})\mathbf{E} = i\omega \mathbf{J}(\mathbf{x}_k), \forall k \end{aligned} \qquad (1)$$

where $L$ is the loss function of the training process, and $\mathbf{T} = (\mathbf{E}_1 \ \mathbf{E}_2 \ \dots \ \mathbf{E}_n)$ is a tensor that contains all the electric field responses with samples in the training dataset. $\mathbf{r}$ is the parameter to be optimized which represents the radii of the holes in the nanophotonic media to be designed. $\mathbf{X}$ and $\mathbf{t}$ are features and labels of the training dataset. $g$ is a function to apply fabrication constraints for each radius and it is dynamically changed with respect to the nanophotonic media distribution. The second constraint is the physical constraint from the Maxwell Equations for the integrated optical structures with silicon and silica, where $\mu_0$ is the permeability of vacuum, $\mathbf{E}$ is the electric field with the dielectric constant $\varepsilon(\mathbf{r})$ and the optical source distribution $\mathbf{J}(\mathbf{x}_k)$ at an angular frequency of $\omega$. The optical source distribution $\mathbf{J}(\mathbf{x}_k)$ is related to $\mathbf{x}_k$ which is the feature of a single sample in the training dataset.

We designed a gradient descent method with projection operations to solve this optimization problem. A schematic of the training process is shown in Fig. **2a**. For each training sample, input sources on the left waveguides are reconfigured for simulation, and the resulting power intensities in the right waveguides are monitored and used to calculate the loss via a loss function. Gradients of the loss function with respect to the hole radii are then computed using the backpropagation method. The nanophotonic media is subsequently updated based on the gradient information and projection operations, incorporating fabrication constraints.

Designing functional structures by solving optimization problems is classified as inverse design problems in integrated photonics, where ensuring the feasibility of fabricating the designed structures presents a significant challenge. An intuitive approach for ensuring fabricability involves selecting a design space that exclusively accommodates discrete, manufacturable geometries. Algorithms such as the direct binary search (DBS) algorithm[20] can be employed to explore this limited discrete design space. While this method consistently produces reliable results, it may restrict the degrees of freedom available for potential designs. Another approach for ensuring fabricability involves selecting a larger, continuous design space that includes non-fabricable geometries. Constraints are then applied within the algorithms to guide the inverse-designed structures away from these non-manufacturable geometries. Techniques like topology optimization[21], shape optimization[22], and methods integrated with the level-set method[23] are successful and elegant for navigating continuous design spaces. In topology optimization, projection methods[24] and morphological operations[25] are commonly used to address non-manufacturability issues. By translating fabrication constraints into penalty terms within the optimization objectives, some methods integrated with the level-set approach have been developed to limit the minimum feature size and minimum radius of curvature[26,27]. Nonetheless, due to the non-convexity of inverse design problems and the discrete nature of the manufacturable design space, incorporating efficient fabrication constraints into inverse design methods remains a significant challenge.

In our approach, the design space is discretized into nano-block arrays while maintaining the radius of the hole structures within each block as continuously tunable, thereby achieving a large design parameter space. At the same time, we specify dynamically changing constraint rules to ensure that all hole structures meet fabrication constraints. After calculating the gradient information using the adjoint method[22] in each iteration, we apply a projection operation based on the gradient information and the current structural state to optimize all structures under fabrication constraints until the iteration limit is reached (details and comparison with topology optimization are described in Supplementary Information S1), as shown in Fig. **2b**. The gradient information is obtained using samples and labels from the task dataset, enabling a training process for the structure to learn and adapt to the target task.

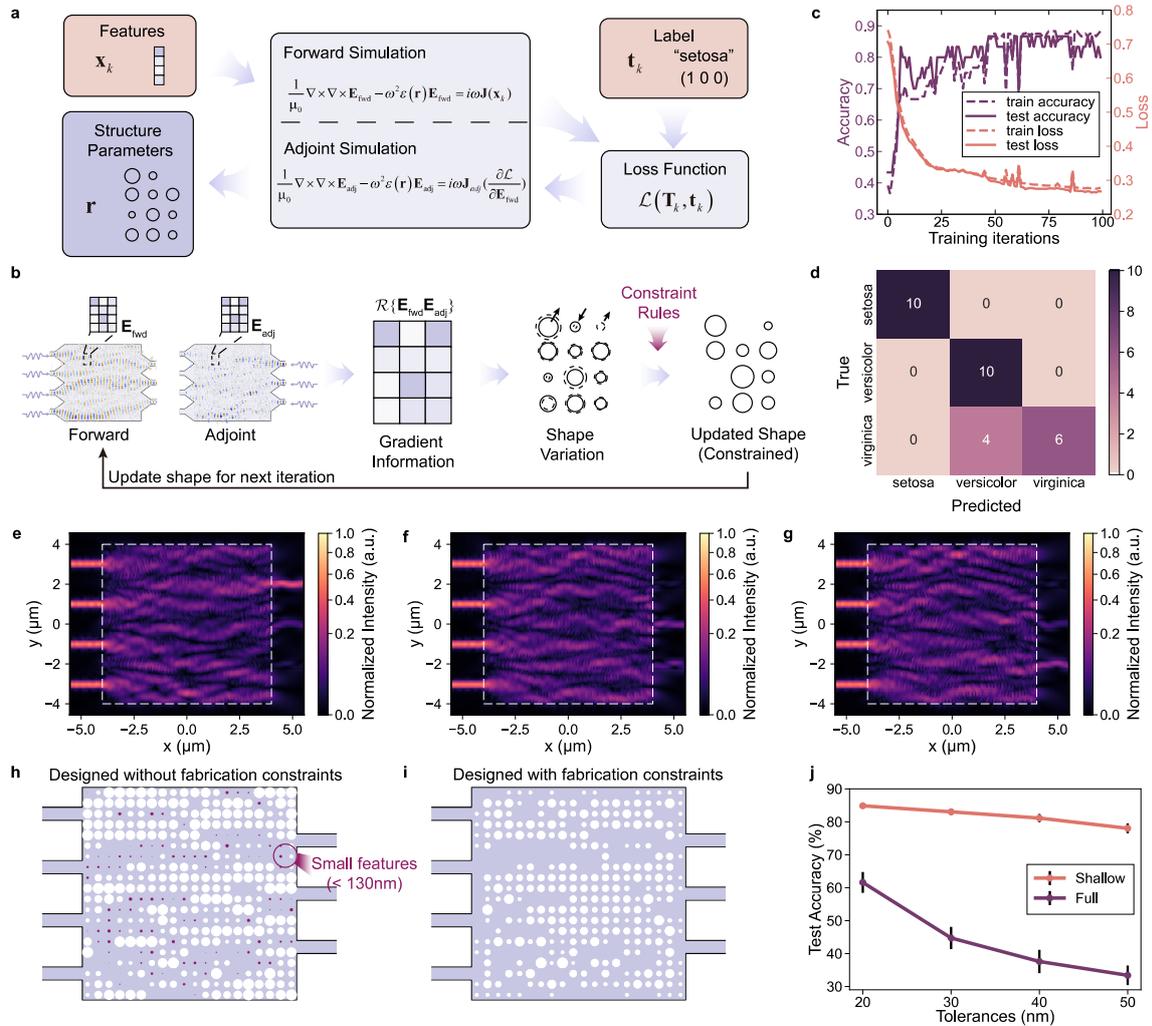

**Fig. 2 | Design flow and simulation results of Iris flower classification. a,** design flow of the structure parameters of the nanophotonic media with the features and label in the training dataset. **b,** structure parameters updating rules in each iteration. **c,** prediction accuracy and cross-entropy loss over 40 training iterations. **d,** confusion matrix for test data. **e-g,** the absolute value of the Poynting vector in x-direction for randomly selected inputs: **e,** prediction of setosa; **f,** prediction of versicolor; **g,** prediction of virginica, a.u. is the abbreviation of arbitrary unit. **h, i,** comparison of designs with (**h**) no fabrication constraints and (**i**) fabrication constraints incorporated

into the design process. **j**, comparison of the simulated fabrication variation under different fabrication tolerances with shallow etching and full etching, where vertical lines represent the 95% confidence interval.

We use the Iris flower dataset to train and test the nanophotonic media. This dataset consists of 150 data points[17] for classifying three types of Iris flowers: setosa, versicolor, and virginica. Each of the data points has four flower features. These features are normalized and rescaled to fit within the 0-π phase range of the eigenmode sources of the input waveguides. The nanophotonic media system is trained using 80% of the dataset, with the remaining 20% reserved for testing the trained system. The nanophotonic media that works as Iris flower classification inference function is an 8×8 µm$^2$ region in the center with 20×20 70 nm shallow etched holes, which is shown in Fig. **1b**. Diameters of these holes vary from 130 nm to 400 nm, ensuring the fabricability. As part of the SOI platform, the 220 nm thick silicon material is on top of a 2 µm silicon oxide box and covered by 1 µm silicon oxide. Diameters of the holes are learned from the Iris flower dataset. After fabrication, the nanophotonic media can identify the category of an Iris flower with unlearned features.

A normalized mean square error (NMSE) loss function (details in Supplementary Information S2) and a batch gradient descent strategy using the adaptive moment estimation (Adam) optimization algorithm are employed for the training process. Fig. **2c** and Fig. **2d** illustrate the evolution of the NMSE loss, prediction accuracy, and the confusion matrix for the test data after training, respectively. An accuracy of 86.7% is achieved for the test data after 95 training iterations. Fig. **2e-g** display the absolute value of the Poynting vector in the x-direction for randomly selected inputs corresponding to the three types of Iris flowers.

After training, the nanophotonic media is capable of processing machine learning tasks with computation performed through scattering blocks composed of numerous holes within the medium. The minimum feature size for the etching process is 130 nm. A comparison of the design without fabrication constraints is shown in Fig. **2h** and Fig. **2i**, where the proposed fabrication constraints not only eliminate excessively small patterns but also facilitate passing the design rule check (DRC) for commercial lithography (details in Supplementary S3). Besides designing for fabrication rule compliance and ensuring fabrication performance, we also significantly enhance the fabrication tolerance of our design by adopting a low-index-contrast approach. A comparison of simulated fabrication variations across different tolerance levels is shown in Fig. **2j**. Fabrication tolerances were analyzed using the Monte Carlo method (see Supplementary S4 for details). Our results indicate that employing low-index-contrast structures effectively reduces the impact of fabrication variations. While the original full-etching design exhibits substantial performance degradation at 20 nm process variations and becomes nearly non-functional at 50 nm variations, the shallow-etch design maintains satisfactory performance even under 50 nm variations. It is worth noting that the 20 nm tolerance value, which serves as the starting point in Fig. **2j**, already represents a conservative estimate, as it exceeds the typical requirements of most real-world fabrication processes.

**Experiment**

The experiment setup with the fabricated chip is shown in Fig. **3a.** The chip is fabricated using electron-beam lithography at the Center for Advanced Electronic Materials and Devices (AEMD) of Shanghai Jiao Tong University. Optical microscopy and scanning electron microscopy images of the nanophotonic media region are shown in Fig. **3b** and Fig. **3c**, respectively. The input light with a wavelength of 1550 nm is coupled into the chip through a single grating coupler and then split by three power splitters to generate four coherent light sources for the nanophotonic media. Four phase modulators are integrated on the waveguides of these four coherent light sources, each capable of providing a phase shift from 0-π with different voltages. To minimize additional phase difference errors, the lengths of the four input waveguides are designed to be equal. At the output of the nanophotonic media, three waveguides and three couplers are used to direct the light carrying the results to output fibers, where it can be detected by off-chip photodetectors. Both simulation and measurement results consistently show an insertion loss of approximately 10 dB (see Supplementary S5 for details), which could potentially be improved by incorporating transmission-enhancing terms into the loss function during the device design optimization process.

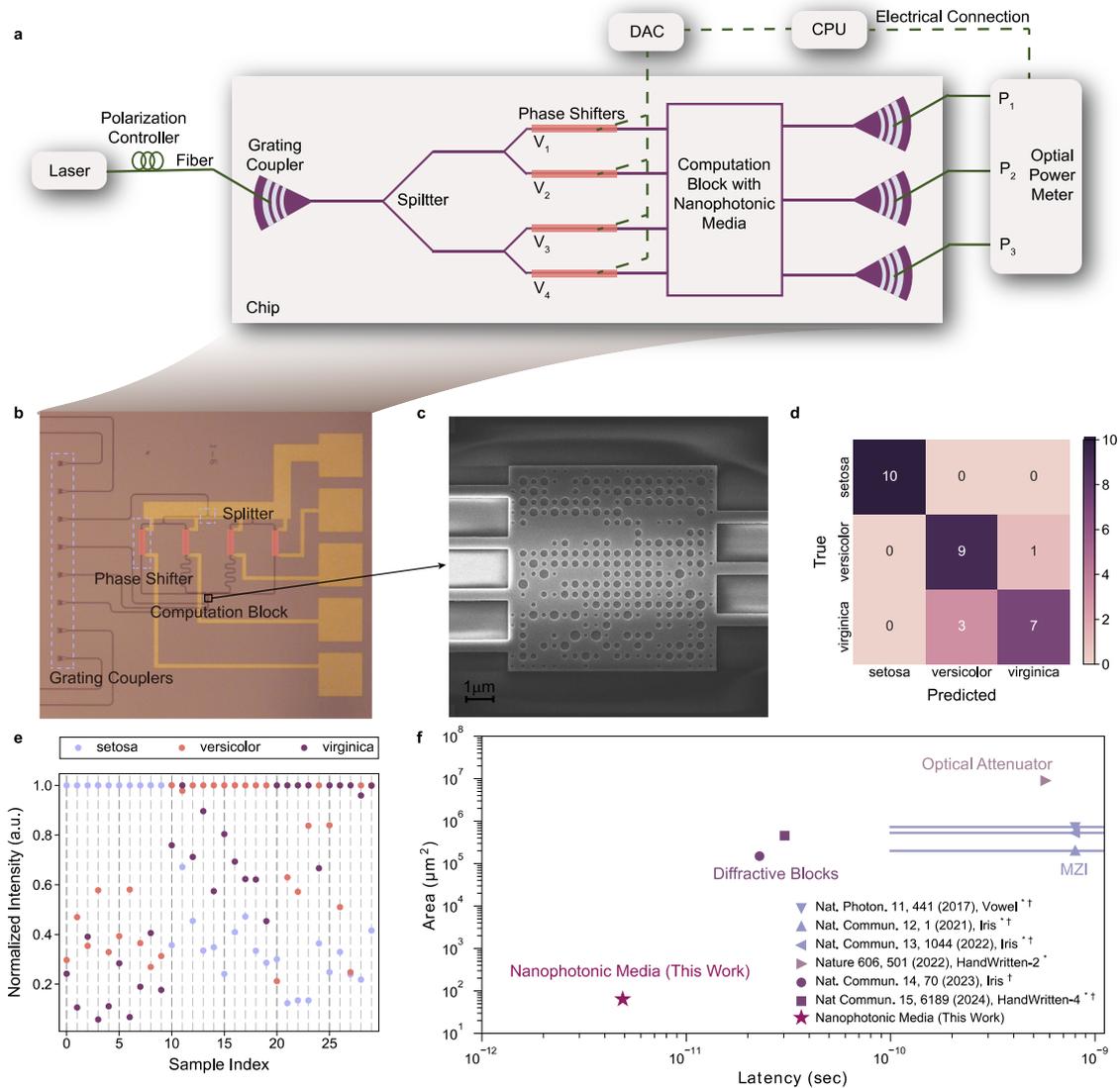

**Fig. 3 | Experiment setup and test results.** **a,** experiment setup for the fabricated chip. **b,** Microphotograph of fabricated chip and **c,** scanning electron microscopy photo for the nanophotonic media. **d,** confusion matrix on the test dataset after compensation. **e,** experimental normalized intensity distributions for all the samples in the test dataset after compensation. The ground truth labels for sample indices 0-9 are setosa, for 10-19 are versicolor, and for 20-29 are virginica. **f,** Comparison of size and latency for on-chip inference systems[4,5,7,11,12,28]. Latency in this work is calculated in Supplementary Information S9. Lines indicate achievable latency ranges through component refinement. HandWritten-n represents the n-category handwritten digits dataset. *Area estimated from microscope images. †Latency obtained through theoretical calculations.

After performing a phase-to-on-chip phase conversion (details are described in Supplementary Information S6), we map the input features to the voltages required for the phase shifters. For each test sample in the dataset, we apply the corresponding voltages to the phase shifters and measure the light power intensities of the three outputs, which represent the probability of the predicted category. We made an adjustment to account for fabrication imperfections and inconsistencies in the grating couplers (details are described in Supplementary Information S7). Prior to this adjustment, the experimental test accuracy was 50.0%. After compensation, the accuracy improved to 86.7%. The confusion matrix after compensation is shown in Fig. **3d** (the one before compensation is shown in Supplementary Information S7). Experimental normalized intensity distributions for all the samples in the test dataset are shown in Fig. **3e**. The experimental inference results are consistent with the designed inference outcomes (simulated normalized intensity distributions are shown in Supplementary Information S7). The broadband capability of the design extends up to 100 GHz, characterized by evaluating frequency shifts of ±50 GHz and ±100 GHz, which correspond to practical modulation scenarios (see Supplementary Information S8 for details).

Fig. **3f** illustrates the comparison of size and latency for on-chip inference systems. The scattering process in this work efficiently mixes and modulates the input light over a short distance, leading to a reduction of more than three orders of magnitude in size for on-chip machine learning inference platforms. This miniaturization also translates to a shorter propagation distance, thereby reducing system latency (latency calculations are provided in Supplementary Information S9). However, compared to the latency introduced by the input section of the system, the latency in the computational block is significantly lower. As a result, Fig. **3f** does not fully reflect the advantage of reduced computational latency. To address this, we calculate the latency of the computational block separately and discuss it further in the Discussion section.

**Hand Written Digit Images Recognition**

To further validate the performance of the proposed nanophotonic media for machine learning inference, a larger region with 64 input waveguides is trained to recognize hand written digit images for optical character recognition (OCR) systems. The open-source dataset[29] comprises 3,823 training images and 1,797 test images. The 8×8 pixels of the images are directly encoded into the mode phase of 64 input waveguides. The nanophotonic media for the optical character recognition as shown in Fig. **4a** consists of a 112×112 array of holes within a 44.8×44.8 μm$^2$ area. NMSE loss and Adam optimization algorithm are adapted for the training process, and a batch gradient descent strategy is employed. Similar to the Iris flower classification task, we employed a low-index-contrast design with a minimum feature size of 130 nm.

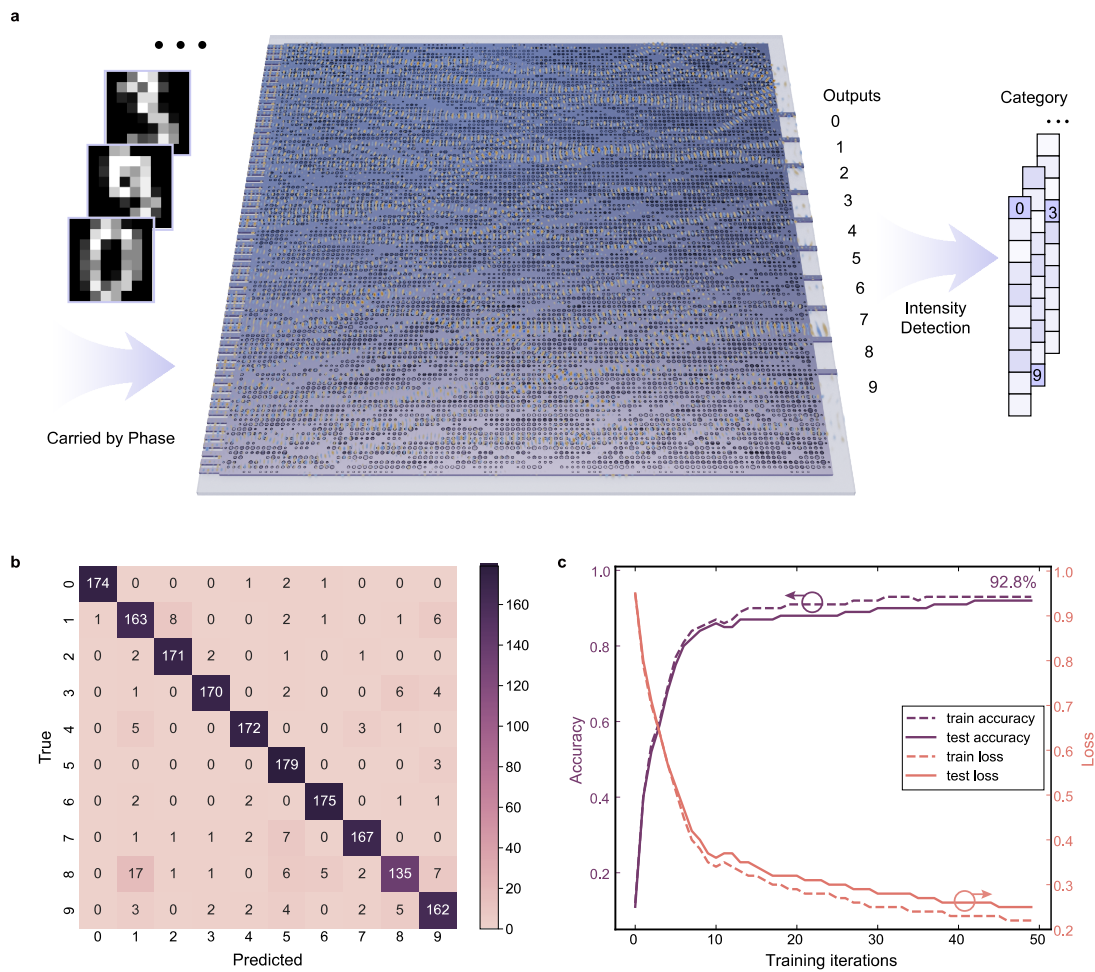

**Fig. 4 | Hand written digit images recognition inference with nanophotonic media. a,** Schematic of hand written digit images recognition inference. **b,** Confusion matrix on test data. **c,** Prediction accuracy and loss over 50 training iterations.

After training, the inference accuracy on the test dataset is 92.8%. Fig. **4b** displays the confusion matrix for the test data after training. Fig. **4c** shows the evolution of the prediction accuracy and NMSE loss in the training process. The final structure is shown in Supplementary S10. The results demonstrate that nanophotonic media have the potential to process large-scale tasks within a very small area, significantly enhancing space efficiency for on-chip optical computations.

**Discussion**.

Our results demonstrate a significant advancement in on-chip machine learning inference through the use of nanophotonic media. Unlike conventional optical neural network designs that rely on systematically structured components, such as Mach-Zehnder interferometers and diffractive elements, our approach leverages a passive photonic medium with a scattering-based architecture, yielding ultra-low power consumption and an ultra-compact footprint. In the experiment, the input optical power is only 1 mW at 1550 nm, along with a peak power consumption of 56 mW for all phase shifters. This demonstrates the energy efficiency of our design, which is on par with or even surpasses some existing on-chip diffractive optical neural networks[11] , known for their high energy efficiency.

Computational density is a fundamental metric for assessing the performance of computational systems, typically quantified by the number of operations executed per unit area in optical computing systems. However, due to the inherent complexities in precisely defining the structural requirements for a single operation within nanophotonic media[13,16], we adopt the approximation that the number of operations necessary for a given computational task remains relatively constant. Consequently, we assess the proposed computing system by evaluating the required structural size and comparing it with other established architectures. To facilitate this comparison, we utilize the Iris flower classification task, a standard benchmark in optical machine learning inference. In parallel, the metric of computational power—expressed as operations per second (OPS)—is determined by the time required to complete the same task. Table 1 provides a comprehensive comparison with other on-chip optical machine learning inference architectures for the Iris flower classification. Our approach achieves an exceptional area reduction of over three orders of magnitude for the same inference task, leading to a corresponding increase in computational density by more than three orders of magnitude. This breakthrough is enabled by a refinement of the core structural elements governing on-chip light propagation, grounded in first-principles design, which allows for ultra-dense integration while preserving high classification accuracy. In terms of computational power, the latency of the computational block, as outlined in Supplementary Information S9, also demonstrates a reduction exceeding one order of magnitude.

Nonlinearity poses a challenge for on-chip machine learning inference. One feasible solution is to use optical-electrical-optical conversion to introduce nonlinearity in the electrical part of the system. However, this approach increases power consumption and system latency. An alternative approach is to exploit the intrinsic nonlinearity of the chip material, although this remains a persistent challenge. Nanophotonic media can address these issues by achieving nonlinear functions in the linear coherent systems[31] (see Supplementary S11 for detailed strategy) or by using highly nonlinear materials as part of the nanophotonic media[32,33].

Although on-chip optical machine learning inference systems are inherently analog and therefore prone to higher errors due to fabrication imperfections, these errors can be substantially reduced by refining the fabrication process and implementing post-fabrication compensation. Additionally, quantization is a common technique used in digital systems to accelerate large-scale machine learning inference by sacrificing some precision. Similarly, on-chip optical machine learning inference can achieve significant power consumption reductions with a trade-off in precision.

In summary, this work designs and experimentally demonstrates a high computational density on-chip optical architecture using nanophotonic media for machine learning inference. The system's ultra-compact size and low power consumption underscore its potential for high-density integration in on-chip machine learning inference systems, offering a promising alternative for complex tasks traditionally handled by diffractive optical neural networks[34,35]. To address the challenges posed by fabrication-induced perturbations in such small areas, fabrication constraints and a low-index-contrast approach are integrated into the design, enhancing the system's tolerance to fabrication errors. Furthermore, its high energy efficiency makes it an excellent candidate for next-generation AI edge computing applications, paving the way for ultra-dense integration and significant energy savings.

## Methods

### Chip Fabrication

The chip is fabricated on a silicon-on-insulator (SOI) wafer with a 220 nm top silicon layer over a 2 µm buried silicon dioxide (SiO2) layer. The silicon waveguides and nanophotonic elements are first patterned using electron beam lithography (EBL) and then fully etched through a single-step inductively coupled plasma dry etching process. Next, the grating patterns and other nanostructures requiring 70 nm shallow etching are patterned and etched. Subsequently, a 1 µm top SiO2 passivation layer is deposited using plasma-enhanced chemical vapor deposition (PECVD). A titanium metal heater with a thickness of 200 nm and a gold metal interconnection also with a thickness of 200 nm are then defined using EBL and sequentially deposited via an electron beam evaporator.

### Experiments

A tunable continuous wave laser (Santec TSL 770) and a power monitor (Santec MPM 210) are utilized for launching input and monitoring output light. A multi-channel voltage-stabilizing source (T2-MS64-5CV) is used for applying voltages for phase shifters. A source meter (Keithley 2400) is used for evaluating the power consumption of phase shifters.

**Simulations**

An FDTD method (https://www.ansys.com/products/optics/fdtd) was used to simulate the field distribution on the nanophotonic media. The 3-dimensional FDTD is adopted for the iris classification task. By calculating the effective index of the silicon slab and resetting the material refractive index, we use the 2-dimensional FDTD as a 2.5D FDTD method for the optical character recognition task. Gradient calculation is implemented with NumPy, which is a package for Python, and the Python interface of Ansys Lumerical FDTD.

**Acknowledgments**

This work was financially supported by the National Key R&D Program of China (2023YFB2804702); Natural Science Foundation of China (NSFC) (62175151, 62341508, 62422509); Shanghai Municipal Science and Technology Major Project. We also thank the Center for Advanced Electronic Materials and Devices (AEMD) of Shanghai Jiao Tong University (SJTU) for fabrication support.

**Author contributions**

X.H.G initiated the project. Z.Y.Z, Y.C.P performed the calculation and simulation. Z.Y.Z and X.H.G. designed the experiments. Z.Y.Z and Y.L.C fabricated samples. Z.Y.Z and Y.H. carried out the measurements. Z.Y.Z, Y.C.P, J.L.X, Y.J.Z, A.H, Y.T.Z, Y.L.C, Y.H, X.Y.F, Y.K.S, M.G and X.H.G analyzed the results and wrote the manuscript. X.Y.F, Y.K.S, M.G and X.H.G supervised the project.

**Conflict of interests**

The authors declare no competing interests.

**Correspondence and requests for materials** should be addressed to Xuhan Guo.

**Data availability**

The data that support the plots within this paper are available from the corresponding authors upon request.

**Table 1 | Comparison of the on-chip machine learning inference with Iris flower dataset**

| Resource | Structures | Area (mm$^2$) | Computational Latency (ps) | With O-E-O Nonlinearity | Experimental Accuracy |
|---|---|---|---|---|---|
| Ref. 5 | MZIs | >0.15 | >4.73 | Yes | 97.4% |
| Ref. 28 | MZIs & Diffractive Blocks | >0.15 | >4.73 | Yes | 96.7% |
| Ref. 11 | SWUs & Diffractive Blocks | 0.15 | 4.73 | No | 86.7% |
| Our work | Nanophotonic Media | 0.000064 | 0.11 | No | 86.7% |

In table 1, MZI Mach–Zehnder interferometer, SWU Subwavelength unit, O-E-O Optical-electrical-optical. N/A indicates no available data.